\documentclass[conference]{IEEEtran}
\IEEEoverridecommandlockouts
\usepackage[space]{cite}
\usepackage{amsmath}
\usepackage{amssymb}
\usepackage{amsfonts}
\usepackage{algorithm}
\usepackage[noend]{algpseudocode}
\usepackage{graphicx}
\usepackage{subfig}
\usepackage{booktabs}
\usepackage{textcomp}
\usepackage{xcolor}
\usepackage{float}
\usepackage{threeparttable}
\usepackage{colortbl}
\usepackage{setspace}
\usepackage{bm}
\usepackage{siunitx}
\usepackage{hyperref}
\hypersetup{colorlinks=true,linkcolor=blue,filecolor=magenta,urlcolor=cyan,citecolor=black}

\newtheorem{definition}{Definition}

\newcommand{\supp}{\operatorname{supp}}

\newcommand{\nei}{\operatorname{Nei}}

\newcommand{\mc}{\mathcal}

\newcommand{\pluseq}{\mathrel{{+}{=}}}

\newcommand{\R}{\mathbb R}
\newcommand{\G}{\mathbb G}

\DeclareMathOperator*{\median}{median}

\def\BibTeX{{\rm B\kern-.05em{\sc i\kern-.025em b}\kern-.08em
    T\kern-.1667em\lower.7ex\hbox{E}\kern-.125emX}}
\begin{document}

\title{\textsc{SubAnom}: Efficient Subgraph Anomaly Detection Framework over Dynamic Graphs}

%\iffalse
\author{\IEEEauthorblockN{Chi Zhang}
\IEEEauthorblockA{\textit{Fudan University} \\
\textit{Shanghai Key Laboratory of Data Science}\\
Shanghai, China \\
21210980092@m.fudan.edu.cn}
\and
\IEEEauthorblockN{Wenkai Xiang}
\IEEEauthorblockA{\textit{Fudan University} \\
\textit{Shanghai Key Laboratory of Data Science}\\
Shanghai, China \\
21210980078@m.fudan.edu.cn}
\and
\IEEEauthorblockN{Xingzhi Guo}
\IEEEauthorblockA{\textit{Stony Brook University} \\
Stony Brook, USA \\
xingzguo@cs.stonybrook.edu}\and
\IEEEauthorblockN{Baojian Zhou\textsuperscript{*}\thanks{\textsuperscript{*}Corresponding Author}}
\IEEEauthorblockA{\textit{Fudan University} \\
\textit{Shanghai Key Laboratory of Data Science}\\
Shanghai, China \\
bjzhou@fudan.edu.cn}
\and
\IEEEauthorblockN{Deqing Yang}
\IEEEauthorblockA{\textit{Fudan University} \\
\textit{Shanghai Key Laboratory of Data Science}\\
Shanghai, China \\
yangdeqing@fudan.edu.cn}
}
%\fi
\vspace{-5mm}
\maketitle
\vspace{-6mm}
\begin{abstract}

Given a dynamic graph, the efficient tracking of anomalous subgraphs via their node embeddings poses a significant challenge. Addressing this issue necessitates an effective scoring mechanism and an innovative anomalous subgraph strategy. Existing methods predominantly focus on designing scoring strategies or employing graph structures that consider nodes in isolation, resulting in ineffective capture of the anomalous subgraph structure information.

In this paper, we introduce \textsc{SubAnom}, a novel framework for subgraph anomaly detection that is adept at identifying anomalous subgraphs. \textsc{SubAnom} has three key components: 1) We implement current state-of-the-art dynamic embedding methods to efficiently calculate node embeddings, thereby capturing all node-level anomalies successfully; 2) We devise novel subgraph identification strategies, which include $k$-hop and triadic-closure. These strategies form the crucial component that can proficiently differentiate between strong and weak neighbors, thus effectively capturing the anomaly structure information; 3) For qualifying the anomaly subgraphs, we propose using $\ell_p$-norm-based score aggregation functions. These iterative steps enable us to process large-scale dynamic graphs effectively.

Experiments conducted on a real-world dynamic graph underscore the efficacy of our framework in detecting anomalous subgraphs, outperforming state-of-the-art methods. Experimental results further signify that our framework is a potent tool for identifying anomalous subgraphs in real-world scenarios. For instance, the F1 score under the optimal subgraph identification strategy, can peak at 0.6679, while the highest achievable score using the corresponding baseline method is 0.5677.

\end{abstract}


\section{Introduction}
\label{sect:introduction}

\begin{figure}[t]
\centering
\includegraphics[width=.5\textwidth]{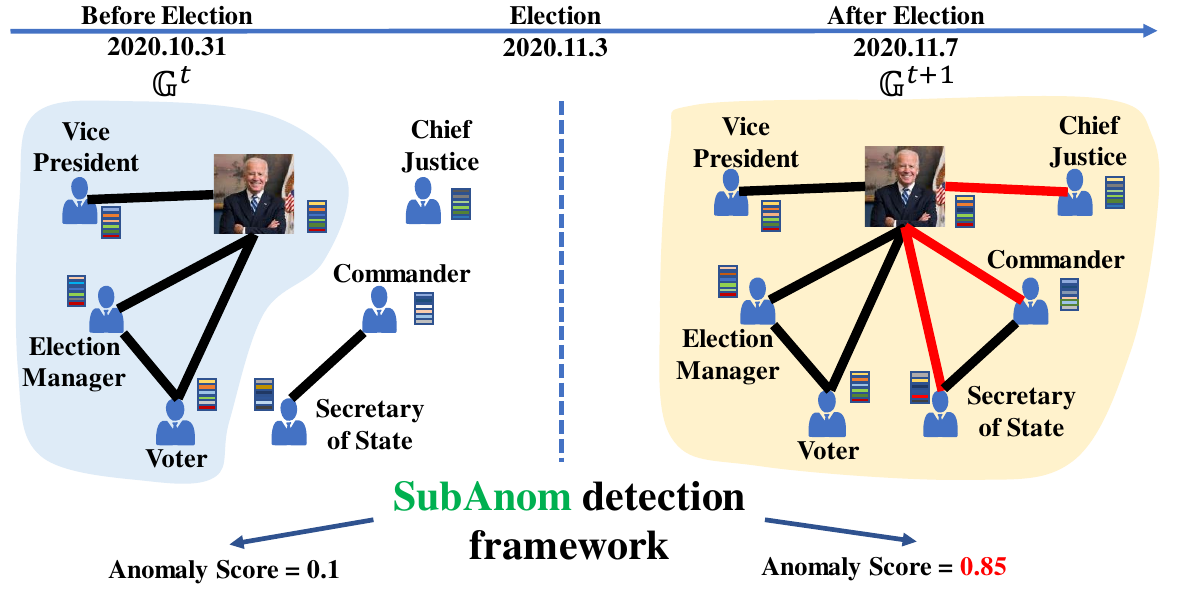}
\caption{The illustration of the \textsc{SubAnom} framework: monitoring the social status change of Joe Biden. The left part exhibits the social graph pre-election, highlighting a subgraph associated with Joe Biden before the 2020 Election. The right part depicts an anomalous subgraph post-election, representing the transition in his social status following Election Day. \textsc{SubAnom} can identify the anomalous subgraphs.\vspace{-5mm}}
\label{fig:intro-toy-example}
\end{figure}

Given a dynamic graph, the task of identifying anomalous subgraphs is of critical importance, with wide-ranging applications such as detecting events in social networks \cite{rozenshtein2014event}, pinpointing anomalous user groups in financial networks \cite{perozzi2016scalable}, unearthing anomalous clusters of reviewers from user-item graphs \cite{hooi2016fraudar}, and many others \cite{miller2010subgraph,miller2015spectral,guo2022subset}. As illustrated in Fig. \ref{fig:intro-toy-example}, a representative example of identifying anomalous subgraphs over dynamic graphs is as follows: take into account the changes in Joe Biden's social status during the 2020 election period. His social status underwent significant alterations, as seen from pre-election to post-election, where additional edges have been integrated, indicating the new social status change. The pivotal aspect of identifying this shift in social status is effectively detecting the anomalous subgraph associated with this transformation. Inspired by this, we pose the following question: \textit{What is the most efficient and effective methodology for identifying anomalous subgraphs over a potentially large-scale dynamic graph?}

Existing methods for subgraph anomalous detection falls into two main categories, including structured-based methods and representation learning-based methods. The methods of structured-based use more traditional algorithms, which involve identifying subgraphs using a user-specified score function. However, one needs to optimize the objective function using complicated optimization algorithms \cite{rozenshtein2014event,zhou2016graph,chen2017generic}. Such methods not only consume a considerable amount of time but also pose challenges when applied to dynamic graphs. In the dynamic setting, one typically has to deal with thousands of snapshots with large size of nodes, making this type of traditional approach less feasible. Thanks to the recent advances in representation learning over dynamic graphs \cite{guo2021subset,guo2022subset,deng2023accelerating}, representation learning-based methods use node embeddings to capture anomalous information. These advances prove that one can efficiently track anomalous nodes via node embeddings. However, existing methods of this type by seeking out anomalous nodes without taking subgraph information into account, as demonstrated in \cite{guo2022subset}, thereby overlooking hidden but crucial anomaly structural information. A natural question is: \textit{How to detect anomalous subgraph structure information using dynamic node embeddings?}

In response to the aforementioned questions, our paper aims to delve into representation learning-based methods, investigating how anomalous subgraphs can be effectively detected through dynamic node embeddings. We introduce a novel framework for anomalous subgraph detection, namely \textsc{SubAnom}, which leverages the power of dynamic graph representation learning algorithms. \textsc{SubAnom} has three critical components: 1) An efficient learning algorithm designed to learn dynamic node embeddings; 2) Novel subgraph identification methods that leverage $k$-hop and triadic-closure principles generated from seeding nodes; and 3) Score aggregation strategies employing $\ell_p$-norm-based score functions. Our contributions are summarized as follows:

\begin{itemize}
\item We introduce \textsc{SubAnom}, an efficient framework for anomalous subgraph detection. It accepts state-of-the-art efficient node embeddings as input and effectively identifies anomalous subgraphs across graph snapshots. It can be readily extended to attributed networks.
\item \textsc{SubAnom} integrates two novel components: 1) It implements simple yet effective subgraph identification strategies, including $k$-hop and triadic-closure methods. These strategies successfully reveal the underlying subgraphs associated with anomalous events. 2) We introduce a range of effective score aggregations that can accurately quantify the anomaly level based on the $\ell_p$-norm.
\item We evaluate \textsc{SubAnom} through experiments on real-world dynamic anomalous subgraph detection tasks. The results confirm the effectiveness and efficiency of \textsc{SubAnom}. Compared with state-of-the-art methods, \textsc{SubAnom} yields significantly higher F1 scores, owing to its ability to capture the hidden anomalous structure.
\end{itemize}

The source code is available at {\small \url{https://github.com/Baderlic/SubAnom}}. The rest is organized as follows: In Section \ref{sect:related-work}, we discuss the related work. Our notations and preliminaries are presented in Section \ref{sect:notations}. Section \ref{sect:proposed-algorithm} outlines our proposed \textsc{SubAnom} framework. The experimental results are provided in Section \ref{sect:experiments}. Finally, we draw conclusions in Section \ref{sect:conclusion}.

\section{Related Work}
\label{sect:related-work}

\textit{\textbf{Graph anomaly detection.\quad}} There are two types of anomaly over dynamic graphs, including graph-level and node-level anomaly. We mainly focus on the graph-level anomaly. Graph anomaly refers to sudden changes in the graph structure during its evolution, measuring the difference between consecutive graph snapshots \cite{aggarwal2011outlier,beutel2013copycatch,eswaran2018sedanspot,yoon2019fast,guo2022subset}. The work of \cite{aggarwal2011outlier} used a structural connectivity model to identify anomalous graph snapshots. \cite{shin2017densealert} adopted tensor factorization to detect subgraphs. The work of \cite{yoon2019fast} determined the anomaly nodes using derivatives of the global PageRank Vectors. The most similar work to ours is \textsc{DynAnom} \cite{guo2022subset}, where authors consider embeddings and detect anomalous nodes individually. However, these methods cannot reveal the node's local anomalous changes and cannot identify individual node changes.  There are also other types, including edge anomaly detection \cite{bhatia2020midas,eswaran2018sedanspot,ranshous2016scalable,chang2021f,eswaran2018spotlight}.

\textit{\textbf{Node embeddings on dynamic graphs.\quad}} Node representation learning on dynamic graphs is a powerful tool for graph mining, mapping nodes into lower-dimensional vectors \cite{kazemi2020representation,tian2021self,guo2021subset}. These vectors can be effectively applied to the detection of anomalous subgraphs. Recent advancements detailed in \cite{guo2021subset} demonstrate that it is possible to efficiently maintain dynamic embeddings for a subset of target nodes within dynamic graphs \cite{guo2022subset}, namely \textsc{DynAnom}. Yet, anomalous events on dynamic graphs tend to propagate through a subgraph structure, which means \textsc{DynAnom} may overlook crucial hidden subgraph structural information.

\textit{\textbf{Triadic closure.\quad}} In a trusted network, triadic closures are likely to develop due to a property called transitivity \cite{easley2010networks}. Owing to its powerful expressive capability, the transitivity of triadic closure has been applied in various statistical relational models, such as Markov logic networks \cite{richardson2006markov}, Problog \cite{de2007problog}, probabilistic soft logic \cite{kimmig2012short}, and relational logistic regression \cite{kazemi2014relational}. In this work, we leverage the properties of triadic closure to identify subgraphs because it inherently capture the local network structure.

\section{Notations and Preliminaries}
\label{sect:notations}

We will first introduce the notations and then present the model for the dynamic graph. We will explain how the dynamic graph embedding algorithm, \textsc{DynPPE}, can aid in the identification of anomalous subgraphs.

\subsection{Notations}
Given a weighted graph ${\G} = (\mc V, \mc E, \mc W)$, where $\mc V$ is the set of nodes, $\mc E \subseteq \mc V \times \mc V$ is the set of edges, and $\mc W$ denotes edge weights. The dynamic weighted graph at a specific time $t$ is $\G^t = (\mc V^t, \mc E^t, \mc W^t)$. We will omit $t$ when the context is clear. The $d$-dimensional embedding vector of node $v$ at time $t$ is represented as ${\bm x}_v^t \in \R^d$ where the $i$-th entry of ${\bm x}_v^t$ is $x_v^t(i)$. The support of $\bm x$ is the set of nonzero indices, i.e., $\supp(\bm x) = \{i: x_i \ne 0\}$. The set of $n$ numbers is denoted as $[n]=\{1,2,\ldots,n\}$. We use $\delta_v^t$ to express the anomaly score of node $v$ at time $t$. Given a subset of nodes $\mc S^t \subseteq \mc V^t$, $\G^t(\mc S^t)$ represents the subgraph induced by $\mc S^t$. Moreover, the notation $\G^t(\mc S_v^t)$ designates the subgraph generated by node $v$ at time $t$, following a specific subgraph strategy. We will expound on this in Section \ref{sect:proposed-algorithm}. All neighbors of  $v$ form a set $\nei(v)$. Given nodes $u$ and $v$ with corresponding embeddings $\bm x_u$ and $\bm x_v$, we measure their embedding distance (corresponding the anomalous score) using the $\ell_p$-norm distance, defined as $\text{dist}(\bm x_u, \bm x_v) \triangleq \| \bm x_u - \bm x_v \|_p = \left(\sum_{i=1}^n \left| x_u(i) - x_v(i) \right|^p\right)^{1/p}$. Intuitively, the larger the distance between two consecutive embeddings $\bm x_v^{t+1}$ and $\bm x_v^{t}$, the higher the anomaly score could be. The adjacency matrix of $\G$ is $\bm A$ with degree $\bm D$.

\subsection{Dynamic Graph Model}

To characterize the dynamics of an evolving graph $\G^t$, we follow the work of \cite{guo2022subset} and define the dynamic graph model as a sequence of edge events. Let $\Delta \mc E^t$ be edge events from $\G^t$ to $\G^{t+1}$, $\Delta \mc E^t \triangleq \{(u_0,v_0,\Delta w(u_0,v_0)), \ldots, (u_m,v_m, \Delta w$ $(u_m, v_m)\}$ where $\Delta w(u_m,v_m)$ could be \textsc{Insertion}, \textsc{Deletion}, or \textsc{Weight Updating}. We adopt the \textit{Discrete Time Dynamic Graph model} (DTDG) \cite{kazemi2020representation} as the following.

\begin{definition}[Discrete Time Dynamic Graph]
A discrete-time dynamic graph contains a sequence of snapshots from a dynamic graph sampled at regularly-spaced times. Specifically, a DTDG model is a sequence of snapshots, $\left\{{\G}^1, {\G}^2, \ldots, {\G}^T\right\}$ where ${\G}^t= \left({\mc V}^t, {\mc E}^t, {\mc W}^t\right)$ is the $t$-th graph snapshot.
\end{definition}

\subsection{Dynamic Graph Embeddings}

In the DTDG model $\G^t$, we leverage \textsc{DynPPE} \cite{guo2021subset} to construct dynamic node embeddings. At the heart of \textsc{DynPPE} is the efficient maintenance of personalized PageRank(PPR) vectors, achieved through a dynamic forward push algorithm \cite{zhang2016approximate,postuavaru2020instantembedding}. For $v \in \mc V^t$, its PPR vector on $\G^t$ is $\bm \pi_v^t$ defined as 
\begin{equation}
\bm \pi_v^t = \alpha \left(\bm I - (1-\alpha)\bm A^\top \bm D^{-1}\right)^{-1} \bm e_v,\label{equ:ppv}
\end{equation}
where $\alpha$ is a damping factor and $\bm e_v$ is an indicator vector with entries $e_v(i)=1$ if $i=v$, and 0 otherwise. The entire \textsc{DynPPE} procedure is illustrated in Algorithm \ref{algo:dynanom-node}. To derive an effective dynamic embedding of $v$, we first approximate a dynamic node embedding $\bm p_v^t$ to match $\bm \pi_v^t$ (as in \textsc{IncrementPush} of Line 4). Subsequently, we utilize two hash functions to ``project'' $\bm p_v^t$ into a lower-dimensional space, thus forming an approximate embedding of $v$, denoted as $\bm x_v$ (as per \textsc{DynNodeRep} of Line 20). 

More specifically, at the beginning of the \textsc{IncrementPush} method, it uses the initial graph $\G^0$ to obtain $\bm p_v^t$ and $\bm r_v^t$ where $\bm r_v^t$ is the dynamic residual vector (Line 7). After receiving the sequence of snapshots $\Delta \mc E^1,\ldots,\Delta \mc E^T$ as input, it updates $\bm p_v^t$ and $\bm r_v^t$ (Line 9-10) using the following equations:
\begin{align}
p_v(u) &= p_v(u) \frac{\sum_{v \in \operatorname{Nei}(u)} w(u,v)+\Delta w(u,v)}{\sum_{v \in \operatorname{Nei}(u)} w(u,v)}, \label{equ:111} \\
r_v(u) &= r_v(u) -\frac{\Delta w(u,v) p_v(u)}{\alpha \sum_{v \in \operatorname{Nei}(u)} w(u,v)}, \label{equ:222} \\
r_v(v) &= r_v(v) + \frac{(1-\alpha)}{\alpha} \frac{\Delta w(u,v)p_v(u)}{\sum_{v \in \operatorname{Nei}(u)} w(u,v)}. \label{equ:333}
\end{align}
For each specific snapshot, we use the above-mentioned equations to update $\bm p_v^t$ and $\bm r_v^t$ (Line 11). These updates ensure the linear invariant property of $\bm p_v^t$ as demonstrated in \cite{zhang2016approximate,guo2022subset}, meaning that $\bm \pi_v^t$ can still be well approximated by $\bm p_v^t$ in \eqref{equ:ppv}.

Upon acquiring $\bm p_v^{t}$, we ``project'' $\bm p_v^t$ into a lower-dimensional space using \textsc{DynNodeRep} (Line 2). It uses two hash functions to facilitate achieving instant embedding \cite{postuavaru2020instantembedding} ($dim=1024$ in our experiments).

\begin{algorithm}
\caption{$\textsc{DynPPE}({\G}^0, \Delta \mc E^1, \ldots,\Delta \mc E^T, v, \epsilon, \alpha)$ \cite{guo2021subset} }
\begin{algorithmic}[1]
\State $\bm p_{v}^{1:T} = \textsc{IncrementPush}(\G^0, \Delta \mc E^1, \ldots,\Delta \mc E^T, v, \epsilon, \alpha)$
\State $\bm x_{v}^{1:T} = \textsc{DynNodeRep}(\bm p_v^1,\ldots ,\bm p_v^T)$
\State \Return $\bm x_{v}^{1:T}$
\Procedure{IncrementPush}{$\G^0, \Delta \mc E^1,..,\Delta \mc E^T, v, \epsilon, \alpha$}
\State $t=0$
\State $\bm p_v^t = \bm 0$ 
\State $\bm r_v^t = \bm 1_v$
\State $\bm p_v^t, \bm r_v^t=$\textsc{DynamicPush}$(\G^0, \bm p_v^t, \bm r_v^t, \epsilon, \alpha)$
\For{$t = 1,2,\ldots,T$}
    \For{$(i,j,\Delta w(i,j)) \in \Delta \mc E^t$}
        \State Update $p_v^t(i), r_v^t(i), r_v^t(j)$ using \eqref{equ:111},\eqref{equ:222}, and \eqref{equ:333} 
    \EndFor
    \State $\bm p_v^t, \bm r_v^t=$\textsc{DynamicPush}$(\G^t, \bm p_v^t, \bm r_v^t, \epsilon, \alpha)$
\EndFor
\State \Return $\bm p_v^{1:T} =[ \bm p_v^1, ... , \bm p_v^T ]$ 
\EndProcedure
\Procedure{DynamicPush}{$\G^t, \bm p_v, \bm r_v, \epsilon, \alpha$}
\While{exists $u$ such that $| r_v(u)| > \epsilon d(u)$}
\State $p_v(u) \pluseq \alpha r_v(u)$
\For{$i \in \operatorname{Nei}(u)$}
\State $r_v(i) \pluseq \frac{(1-\alpha)  r_v(u)\cdot w(u, v)}{ \sum_{j \in \operatorname{Nei}(u)} w(u, j)} \textcolor{blue}{}$
\EndFor
\State $r_v(u) = 0 $
\EndWhile
\State \Return $(\bm p_v, \bm r_v)$
\EndProcedure
\Procedure{DynNodeRep}{$\bm p_v^1, \ldots ,\bm p_v^T$}
\State $\epsilon_{c}= \textsc{min}(\frac{1}{|\mathcal{V}|}, \text{1e-4}), dim = 1024$
\For{$t = 1,2,\ldots,T-1$}
    \For{$i \in \cup_{t'\in\{t, t-1\}} {\supp}(\bm p_v^{t'})$}:
    \State $p_v^{t-1}(i) = 0$ if $p_v^{t-1}(i) \leq \epsilon_c $
    \State $p_v^t(i) = 0$ if $p_v^t(i) \leq \epsilon_c $
    \EndFor
    \State $\bm x_v^{t} = \textsc{ReduceDim}({\bm p_v^t}, dim)$
    \State \textbf{return} $ \bm x_v =[ \bm x_v^0,\ldots, \bm x_v^T ]$ 
\EndFor
\EndProcedure
\Procedure{ReduceDim}{$\bm x,dim$}
\If{$\textsc{dim}(\bm x) \leq dim  $}
    \State \textbf{return} $\bm x$
\Else
    \State $\bar {\bm x} = \bm 0 \in \R^{dim}$
    \For{$i \in {\supp}(\bm x)$}
        \State $\bar {\bm x}(h_{dim}(i)) \pluseq h_{\operatorname{sgn}}(i) \log \left(x(i) \right)$
    \EndFor
    \State \textbf{return} $\bar {\bm x}$
\EndIf
\EndProcedure
\end{algorithmic}
\label{algo:dynanom-node}
\end{algorithm}
\vspace{-2mm}

\section{Proposed Framework: \textsc{SubAnom}}
\label{sect:proposed-algorithm}

In this section, we present the novel \textsc{SubAnom} framework based on the dynamic node representation learning technique. We first define our problem and then propose to use $k$-hop, triadic-closure, and their hybrid to identify anomalous subgraphs. We then define several anomalous score aggregation strategies to quantify the anomalous subgraphs.

\subsection{Problem Formulation}

This section defines the process of anomalous subgraph detection over dynamic graphs, including subgraph identification and anomaly score aggregation.

\textit{\textbf{Subgraph identification.\quad}} Given the initial graph $\G^0=({\mc V}^0, {\mc E}^0, {\mc W}^0)$, the subgraph identification procedure commences with the identification of an anomalous \textit{seed node} $v\in \mc V^0$. Following this, a \textit{subgraph generation strategy} is implemented, using the selected seed node as a reference to identify a subgraph $\G^0(\mc S_v^0)$ where $\mc S_v^0$ represents the subset of nodes resulting from this generation strategy. We formally define subgraph identification as the following:

\begin{definition}[Subgraph Identification]
Given a dynamic graph $\G^t$ with an initial snapshot $\G^0 = ({\mc V}^0, {\mc E}^0, {\mc W}^0)$, the subgraph identification for a node $v \in \mathcal{V}^t, t=0,1,\ldots,T $, is a strategy designed to find a subset of nodes that forms a subgraph $ \G^t(\mc S_v^t) = \G({\mc V'}, {\mc E'}, \mc W')$ where ${\mc V'}, {\mc E'}, \mc W'$ are subsets of ${\mc V}^t, {\mc E}^t, \mc W^t$, respectively.
\end{definition}

\textit{\textbf{Anomaly score aggregation.\quad}} To determine the anomaly score of the generated subgraph at $t$ given $v$, it is necessary to aggregate the anomaly scores of nodes in $\mc S_v^t$. The definition of anomaly score aggregation is as the following:

\begin{definition}[Anomaly Score Aggregation]
Given a graph $\G = ({\mc V}, {\mc E}, {\mc W})$ and a subgraph $\G(\mc S)$, for every node $v \in {\mc S}$, with $\delta_v$ representing its anomaly score between two consecutive embeddings, the anomaly score of $\G(\mc S)$ is defined as
\begin{equation}
Score_G = f(\delta_v),
\end{equation}
where $f$ is the aggregation function.
\end{definition}
With the above definitions, we are ready to present our proposed generation strategies and score aggregation functions.

\subsection{Anomalous Subgraph Identification}

\begin{figure}[!ht]
\begin{minipage}{0.3\linewidth}
\centerline{\includegraphics[width=3cm]{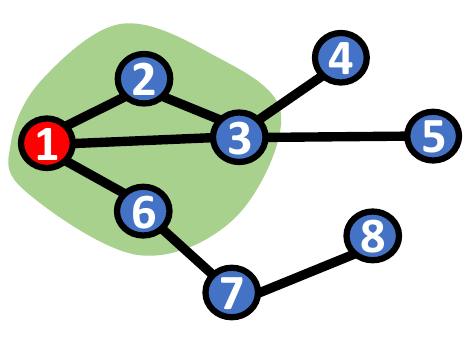}}
\centerline{1-hop subgraph}
\end{minipage}
\qquad\quad
\begin{minipage}{0.3\linewidth}
\centerline{\includegraphics[width=3cm]{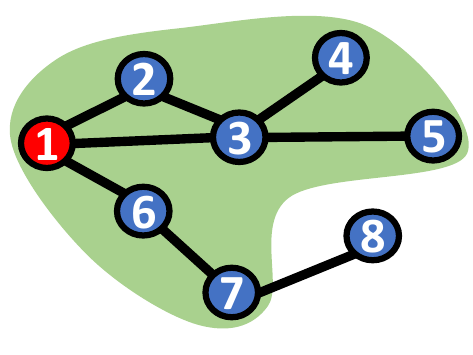}}
\centerline{2-hop subgraph}
\end{minipage}
\centering
\caption{$k$-hop strategy. \textit{Left}: 1-hop $\mc S=\{1,2,3,6\}$ with $v=1$; \textit{Right}: 2-hop $\mc S=\{1,2,3,4,5,6,7\}$. }
\label{fig:k-hop}
\end{figure}

We consider two types of subgraph generation strategies, including $k$-hop and triadic-closure strategies.

\textit{\textbf{$k$-hop identification strategy ($k$-hop).\quad}} Given the seed node $v \in \mc S^t$, one can use Breadth First Search (BFS) starting from $v$ but only exploring $k$-hop neighbors of $v$. For $k$-hop strategies as shown in Fig. \ref{fig:k-hop} and given the seed node $v=1$, the subgraph of selected nodes contains all the nodes within $k=1$ hop and $k=2$ hops of $v=1$. Note that $k$ is usually a small number for anomalous subgraph detection. In our experiments, we adopt three $k$-hop strategies: $1$-hop, $2$-hop, and $3$-hop, which include all the nodes within 1, 2, or 3 hops.

\textit{\textbf{Triadic-Closure strategy (TC).\quad}} We design an effective method using triadic closure \cite{granovetter1973strength}. It is based on the well-known triadic closure property: ``A friend of my friend is my friend,'' a phenomenon widely documented in the literature \cite{yin2019local,yin2020measuring}. \textit{ We posit a similar observation in the context of anomalous: ``An anomalous neighbor of an anomalous neighbor from a seed node is also likely to be anomalous''.}

\begin{figure}[htbp]
\vspace{-4mm}
\begin{minipage}{0.36\linewidth}
\centerline{\includegraphics[width=3cm]{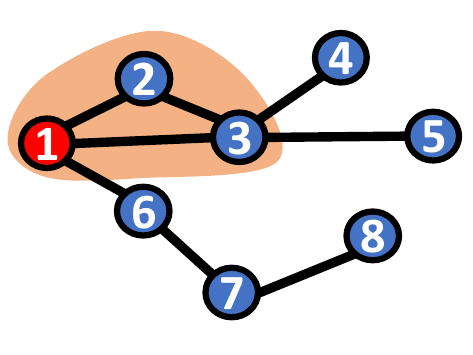}}
\centerline{TC with strong neighbors only}
\end{minipage}
\qquad\quad
\begin{minipage}{0.36\linewidth}
\centerline{\includegraphics[width=3cm]{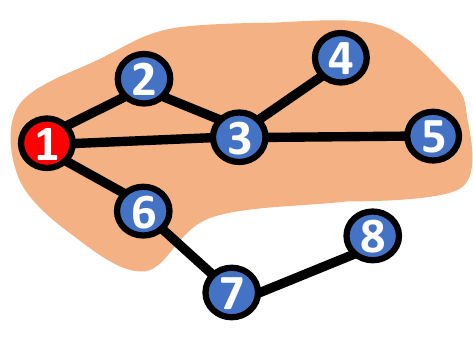}}
\centerline{TC with boundaries}
\end{minipage}
\centering
\caption{Triadic-Closure (TC) strategies. \textit{Left:} TC subgraph with $1$-hop $\mc S=\{1,2,3\}$; \textit{Right:} Hybrid TC subgraph with $1$-hop and boundaries $\mc S = \{1,2,3,4,5,6\}$. \vspace{-5mm}}
\label{fig:triadic}
\end{figure}

To utilize the triadic-closure strategy, we first pinpoint the neighbors of a given node $v$ using the principles of triadic closure. In this context, if node $v$ and two of its neighbors, $u$ and $w$, share direct connections with each other, they form a triadic closure. We then categorize $u$ and $w$ as \textit{strong neighbors} of $v$. Conversely, if $u$ and $w$ lack a direct connection, we label them as \textit{weak neighbors}.

For example, refer to Fig. \ref{fig:triadic} (left). If we use the seed node $v=1$ to generate a $1$-hop subgraph, nodes $2$ and $3$ would be designated as strong neighbors. This is because they each share a direct connection with the seed node and, more importantly, with each other, forming a triadic closure. Thus, by identifying and classifying these strong and weak neighbors, we can incorporate the concept of triadic closure into our framework for anomalous subgraph detection. This strategy provides a more nuanced and detailed view of the graph's structure, which can enhance our ability to detect anomalous patterns.

\begin{table}[]
\small
    %\centering\renewcommand{\arraystretch}{1.5}
    \begin{tabular}{c|c}
    \toprule \rowcolor{gray!40}
       Name of $\phi$ & Aggregation function $\phi$ \\ \hline
        \textsc{Mean} &  $\phi_{\textsc{Mean}}(\bm x_{\cdot}^{t-1:t}, \mc S_v^t) = \frac{1}{|\mc S_v^t|} \sum_{i\in \mc S_v^t} \operatorname{dist}(\bm x_i^{t-1}, \bm x_i^t) $ \\ \hline
        \textsc{Sum} & $\phi_{\textsc{Sum}}(\bm x_{\cdot}^{t-1:t}, \mc S_v^t) = \sum_{i\in \mc S_v^t} \operatorname{dist}(\bm x_i^{t-1}, \bm x_i^t)$ \\ \hline
        \textsc{Max} & $\phi_{\textsc{Max}}(\bm x_{\cdot}^{t-1:t}, \mc S_v^t) =  \max_{i\in \mc S_v^t} \operatorname{dist}(\bm x_i^{t-1}, \bm x_i^t)$ \\ \hline
        \textsc{Min} & $\phi_{\textsc{Min}}(\bm x_{\cdot}^{t-1:t}, \mc S_v^t) =  \min_{i\in \mc S_v^t} \operatorname{dist}(\bm x_i^{t-1}, \bm x_i^t)$ \\ \hline
        \textsc{Median} & $\phi_{\textsc{Median}}(\bm x_{\cdot}^{t-1:t}, \mc S_v^t) = \median_{i\in \mc S_v^t} \operatorname{dist}(\bm x_i^{t-1}, \bm x_i^t)$ \\
        \bottomrule
    \end{tabular}
    \caption{Subgraph aggregation function $\phi$\vspace{-5mm}}
    \label{tab:phi}
\end{table}

\begin{algorithm}[H]
\caption{Hybrid-TC$\left(v, \G^t\right)$}
\label{alg:Hybric-TC}
\begin{algorithmic}[1]
\State \textbf{Input:} Snapshot $\G^t = ({\mc V}^t, {\mc E}^t, \mc W^t)$, node $v \in {\mc V}^t$
\State Construct the adjacency matrix $\bm A^t$ based on $\mc E^t$
\State $\bm C^t = \bm A^t \cdot \bm A^t$
\State Find neighbors of $v$, $\nei(v)$
\State Find closure nodes $w$ such that $\bm C^t[v][w] \neq 0$, $\operatorname{Clo}(v)$
\State Get the strong neighbors of $v$: $\Xi := \nei(v) \cap \operatorname{Clo}(v)$
\State Obtain the boundary nodes $\partial_v $ of $\Xi$;
\State Get all aggregated nodes: $\mc S_v^t = \partial_v \cup \Xi \cup \{ v\}$
\State\Return $\mc S_v^t$
\end{algorithmic}
\end{algorithm}

\textit{\textbf{Hybrid Triadic-Closure strategy (Hybrid TC).\quad}} In the previous triadic-closure strategy could potentially overlook anomalous nodes that, while not classified as strong neighbors, exist on the boundaries of these generated subgraphs. To tackle this limitation, we introduce the Hybrid TC strategy, which includes the boundary nodes of the strong subgraphs as shown in Fig. \ref{fig:triadic} (right), likely to transmit anomalous information emanating from the seed node. This inclusion enables a smoother flow of anomalous information, improving the detection capabilities. To detect triadic closures at $t$, we construct $\bm A^t$ from $\mc E^t$ of $\G^t = ({\mc V}^t, {\mc E}^t, {\mc W}^t)$. For each $v \in \mc V^t$ and its neighbor $u \in \operatorname{Nei}(v)$, we utilize the Common Neighbors Matrix, defined as $\bm C^t = \bm A^t \bm A^t$, to ascertain whether $u$ is a strong neighbor of $v$. If nodes $v$ and $u$ have common neighbors, $\bm C^t[u][v] \neq 0$, they are  strong neighbors.

We elaborate on the Hybrid TC strategy for a given seed node $v$ in the graph $\G = ({\mc V}, {\mc E}, \mc W)$ in Algorithm \ref{alg:Hybric-TC}. Specifically, we first identify neighbors of $v$ and then strong neighbors are identified as the intersection of $\nei(v)$. Boundary nodes of the intersection, $\nei(v) \cap \operatorname{Clo}(v)$, are obtained in Line 7. The final set of nodes under the Hybrid TC strategy is the union of the strong neighbors and these boundary nodes.

\subsection{Anomalous Subgraph Quantification}

Once we have obtained the node embeddings, we need to compute a score to quantify the anomaly of a node and, subsequently, the anomalous subgraph. It has been shown that the $\ell_p$-norm distance between previous and current snapshots can be employed to calculate the anomaly \cite{guo2022subset}. This is because a greater distance signifies, a more pronounced change between the snapshots for a particular node, indicating a higher likelihood of the node being anomalous.  Once a subgraph of a seed node has been generated through a particular strategy, we can aggregate the anomaly scores of the nodes within this subgraph to the seed node. For a node $v$ at the snapshot $t$ associated with an identified subgraph $\G(\mc S_v^t)$, we calculate the aggregate score for all nodes in $ \mc S_v^t$ as follows:
\begin{equation}
SubScore_v^t = \phi(\bm x_{\cdot}^{t-1:t}, \mc S_v^t),
\end{equation}
where $\bm x_{\cdot}^{t-1:t}$ represents graph embeddings of nodes on the subgraph from snapshot $t-1$ to $t$, while $SubScore_v^t$ is the updated anomaly score of node $v$ at snapshot $t$ after graph-level anomaly scores have been aggregated. The aggregation function $\phi$ is applied at the subgraph level and can take the forms of functions shown in Table \ref{tab:phi}.

For graph-level anomaly detection tasks, we also need to aggregate the anomaly scores of all subgraphs to determine the anomaly score for a specific snapshot. For the dynamic graph $\G^t = ({\mc V}^t, {\mc E}^t, \mc W^t)$ and every $v \in {\mc V}^t, t=1,\ldots,T$, we have the final aggregation function
\begin{equation}
Score_{\G}^t =  f\left(\phi\left(\bm x_{\cdot}^{t-1:t},  \mc S_v^t\right)\right), 
\end{equation}
where $f$ is the global aggregation function. The score $Score_G^t$ represents the anomaly score of the graph in snapshot $t$, and $f$ is the graph-level aggregation function. Typically, we can utilize the \textsc{Mean} function for $f$. 

The complete framework is presented in Alg. \ref{alg:SubAnom}. It takes an initial graph $\G^0$, a sequence of edge events forming $T$ snapshots, a precision parameter $\epsilon$, and a teleportation factor $\alpha$ as inputs. For each $\G^t$, it first generates a set of seed nodes $\Theta^t$. For each seed, we calculate the graph embeddings $\bm x_v^{1:t}$. Next, we identify an anomalous subgraph originating from node $v$. We then apply aggregations to compute the subgraph score. After calculating all the subgraph scores, we can finally compute the overall anomaly score $Score_\G^t$.

\begin{algorithm}[H]
\caption{\textsc{SubAnom}($\G^0, \Delta \mc E^{1},\ldots, \Delta \mc E^{T}, \epsilon, \alpha$)}
\label{alg:SubAnom}
\begin{algorithmic}[1]
\State \textbf{Input}: $\G^0$, events $\Delta \mc E^{1},\ldots, \Delta \mc E^{T}$, $\epsilon$, $\alpha$
\For {$t = 1,2,\ldots,T$}
\State Generate a set of seed nodes $\Theta^t$
\For { $v \in \Theta^t$}
\State $\bm x_{v}^{1:t} = \textsc{DynPPE}({\G}^0, \Delta \mc E^1, \ldots,\Delta \mc E^T, v, \epsilon, \alpha)$
\State Generate $\G^t(\mc S_v^t)$ under the subgraph strategy;
\State Calculate the subgraph score $\phi\left(\bm x_{\cdot}^{t-1:t}, \mc S_v^t\right)$
\EndFor
\State $Score_{\G}^t = \frac{1}{|\mc S_v^t|} \sum_{v\in \mc S_v^t}\phi\left(\bm x_{\cdot}^{t-1:t}, \mc S_v^t\right), $
\EndFor
\State \Return $Score_\G^1,Score_\G^2,\ldots,Score_\G^T$
\end{algorithmic}
\end{algorithm}

\subsection{Searching Seed Nodes}

\newcolumntype{?}{!{\vrule width 1pt}}

\begin{table*}[ht]
\centering
        \begin{tabular}{l?llll?llll}
        \toprule\rowcolor{gray!40}
            Methods &  $Score_{\G}^t$ & F1 & Precision & Recall & $Score_{\G}^t$ & F1 & Precision & Recall\\ \hline
            \textsc{SubAnom}(1-hop) & sum-$\ell_1$ & 0.5343 & 0.5760 & 0.4983 & sum-$\ell_2$ & \underline{0.6419} & \underline{0.6920} & \underline{0.5986}\\ \hline
            \textsc{SubAnom}(2-hop) & max-$\ell_1$ & 0.5232 & 0.5640 & 0.4879 & sum-$\ell_2$ & 0.5417 & 0.5840 & 0.5052\\ \hline
            \textsc{SubAnom}(3-hop) & median-$\ell_1$ & 0.5343 & 0.5760 & 0.4983 & sum-$\ell_2$ & 0.5492 & 0.5920 & 0.5121\\ \hline
            \textsc{SubAnom}(1-hop TC) & sum-$\ell_1$ & 0.5195 & 0.5600 & 0.4844 & sum-$\ell_2$ & 0.5380 & 0.5800 & 0.5017\\ \hline
            \textsc{SubAnom}(Hybrid TC) & sum-$\ell_1$ & \underline{0.5380} & \underline{0.5800} & \underline{0.5017} & median-$\ell_2$ & \textbf{0.6679} & \textbf{0.7200} & \textbf{0.6228}\\ \hline
            \textsc{DynAnom} & $\ell_1$ & 0.4675 & 0.5040 & 0.4360 & $\ell_2$ & 0.4935 & 0.5320 & 0.4602 \\ \hline
        \textsc{DynAnom}(High-degree) & $\ell_1$ & \textbf{0.5417} &  \textbf{0.5840} & \textbf{0.5052} & $\ell_2$ & 0.5677 & 0.6120 & 0.5294 \\ \bottomrule
        \end{tabular}
        \caption{The performance of anomalous subgraph detection task using the \textit{weighted graph}. The middle column displays results obtained using the $\ell_1$ distance, while the right column shows results using the $\ell_2$ distance. The values in bold represent the best results, and the underlined values signify the second-best results. The column $Score_{\G}^t$ means the optimal score function.\vspace{-2mm} }
        \label{tab:darpa-da}
    \end{table*}
An essential step of \textsc{SubAnom} is identifying seed nodes. We can identify the top nodes where these embedding distances have changed significantly. However, this approach would still require the computation of dynamic node embeddings for all nodes in $\mc V^t$, which might not be computationally efficient. To bypass such computational redundancies, another possible way is focusing on higher-degree nodes as potential seed nodes, an approach that aligns with the one proposed in \cite{guo2022subset}. High-degree nodes often play critical roles, such as hubs in social networks. Therefore, targeting these nodes could allow us to capture the most salient anomalous behaviors.

\subsection{Complexity Analysis}
The time complexity of \textsc{SubAnom} contains the run time of obtaining dynamic graph embeddings and anomalous subgraph identification. As proved in \cite{guo2022subset}, the total time complexity of dynamic node embedding is nearly-linear to $\Delta m$ where $\Delta m$ is the total number of edge events that happened in all $T$ snapshots. In subgraph identification, the complexity of subgraph identification strategies depends on $\bm A^t \cdot \bm A^t$. Since we use the sparse version of the adjacency matrix $\bm A^t$ to store and search the neighbors of each node, the time complexity is about $\mathcal{O}(|\mc V^t|\cdot {\bar{d}}^2)$ where $\bar{d}$ is the average degree of $\G^t$.  The complexity of $k$-hop strategies is proportional to the size of the nodes explored, usually taking a small portion of the total run time.  The space complexity depends on $\epsilon$. In practice, $\bm p_v^t$ and $\bm r_v^t$ are sparse vectors. Hence, the total space complexity is about $\mc O(n \cdot |\supp(\bm p)|)$ where $|\supp(\bm p)|$ denotes the number of nonzeros of $\bm p$ in expectation.
\section{Experiments}
\label{sect:experiments}

To validate our proposed framework, \textsc{SubAnom}, we perform graph-level anomaly detection on a real-world dataset. Our goal is to address two key questions: \textit{Q1. Does the implementation of a subgraph strategy enhance the detection ability? Q2. Does our method surpass the performance of existing state-of-the-art methods?}

\subsection{Dataset and Baseline Methods}

\textit{\textbf{Dataset.\quad}} DARPA is a dynamic network traffic graph where each node represents an IP address, and each edge signifies network traffic \cite{lippmann20001999}. Network attacks (e.g., DDoS attacks) are manually annotated as anomalous edges. As per common practice in \cite{guo2022subset}, we align the periods of a high anomaly with real-world events. In alignment with the task setting outlined in \cite{guo2022subset,yoon2019fast}, the objective of graph-level anomalous subgraph detection is to compute the anomaly score for $\G^t$. Essentially, we seek the anomalous subgraph, compute its anomaly score, and designate the highest-scoring ones as anomalous snapshots, following the same experimental settings as in \cite{yoon2019fast}. The DARPA dataset consists of $25,525$ nodes, $4,554,344$ edges, $1,463$ snapshots, $256$ initial snapshots, and $289$ anomaly snapshots. We consider both \textit{weighted} and \textit{unweighted}. The goal is to detect 289 anomaly snapshots.

\textit{\textbf{Baseline methods.\quad}} We examine two state-of-the-art methods, including \textsc{DynAnom}\cite{guo2022subset} (designed for dynamic weighted graphs) and \textsc{DynPPE}\cite{guo2021subset} (created for dynamic unweighted graphs where weights are set to unit). It is worth noting that the performance of well-known methods such as \textsc{AnomRank} \cite{yoon2019fast}, \textsc{SedanSpot} \cite{eswaran2018sedanspot} is inferior to that of \textsc{DynAnom}. To compare with previous results, we denote the baselines as \textsc{DynAnom}(High-degree) and \textsc{DynPPE}(High-degree) when the high-degree nodes as seeding nodes, while \textsc{DynAnom} and \textsc{DynPPE} consider all nodes as seeding nodes.

\begin{table*}[ht]
     \centering
        \begin{tabular}{l?llll?llll}
        \toprule\rowcolor{gray!40}
            Methods & $Score_{\G}^t$ & F1 & Precision & Recall & $Score_{\G}^t$ & F1 & Precision & Recall\\ \hline
            \textsc{SubAnom}(1-hop) & sum-$\ell_1$ & 0.4341 & 0.4680 & 0.4048 & sum-$\ell_2$ & 0.4490 & 0.4840 & 0.4187\\ \hline
            \textsc{SubAnom}(2-hop) & min-$\ell_1$ & 0.4267 & 0.4600 & 0.3979 & min-$\ell_2$ & 0.4341 & 0.4680 & 0.4048\\ \hline
            \textsc{SubAnom}(3-hop) & median-$\ell_1$ & \textbf{0.4527} & \textbf{0.4880} & \textbf{0.4221}  & median-$\ell_2$ & \underline{0.4527} & \underline{0.4880} & \underline{0.4221}\\ \hline
            \textsc{SubAnom}(1-hop TC) & sum-$\ell_1$ & 0.4119 & 0.4440 & 0.3841 & sum-$\ell_2$ & 0.4267 & 0.4600 & 0.3979\\ \hline
            \textsc{SubAnom}(Hybrid TC) & sum-$\ell_1$ & \underline{0.4378} & \underline{0.4720} & \underline{0.4083}  & sum-$\ell_2$ & \textbf{0.4601} & \textbf{0.4960} & \textbf{0.4291}\\ \hline
            \textsc{DynPPE} & $\ell_1$ & 0.3673 & 0.3960 & 0.3426 & $\ell_2$ & 0.3785 & 0.4080 & 0.3529\\ \hline
            \textsc{DynPPE}(High-degree) & $\ell_1$ & 0.4341 & 0.4680 & 0.4048 & $\ell_2$ & 0.3859 & 0.4160 & 0.3599\\
            \bottomrule
        \end{tabular}
        \caption{The performance of anomalous subgraph detection task using the \textit{unweighted} graph. The middle column is the results using $\ell_1$ distance while the results in the right column use $\ell_2$ distance. The values in bold represent the best results, and the underlined values signify the second-best results. \vspace{-4mm}}
        \label{tab:darpa-dp}
    \end{table*}

\textit{\textbf{Hyper parameter settings.}} To compare with \textsc{DynAnom} and \textsc{DynPPE}, we adopt the same hyperparameter setting, with $\epsilon = 0.01, \alpha = 0.15,  $ dim$= 1024$. We adopt $f=\textsc{Mean}$ as the final graph aggregation function to compare with baselines.  For the seed node generation at each time $t$, we consider all nodes in $\G^t$ as seed nodes in \textsc{SubAnom}.

\subsection{Comparison of Subgraph Strategies}

\begin{table}[ht]
\centering
    \begin{tabular}{clll}
    \toprule \rowcolor{gray!40}
        Aggregate Method & F1 & Precision & Recall \\ \hline
        Sum-$\ell_1$ & 0.5380 & 0.5800 & 0.5017 \\ \hline
        Sum-$\ell_2$ & \underline{0.6642} & \underline{0.7160} & \underline{0.6194} \\ \hline
        Median-$\ell_1$ & 0.5121 & 0.5520 & 0.4775 \\ \hline 
        Median-$\ell_2$ & \textbf{0.6679} & \textbf{0.7200} & \textbf{0.6228} \\ \hline
        Mean-$\ell_1$ & 0.4972 & 0.5360 & 0.4637 \\ \hline
        Mean-$\ell_2$ & 0.5714 & 0.6160 & 0.5329 \\ \hline
        Max-$\ell_1$ & 0.4972 & 0.5360 & 0.4637 \\ \hline
        Max-$\ell_2$ & 0.5121 & 0.5520 & 0.4775 \\ \hline
        Min-$\ell_1$ & 0.5009 & 0.5400 & 0.4671 \\ \hline
        Min-$\ell_2$ & 0.5083 & 0.5480 & 0.4740\\ \hline
        \textsc{DynAnom-$\ell_1$} & 0.4925 & 0.5280 & 0.4615 \\ \hline \textsc{DynAnom-$\ell_2$} & 0.4478 & 0.4800 & 0.4196\\ \bottomrule
    \end{tabular}
    \caption{The performance of aggregations and distance metrics. Dynamic node embeddings are obtained using the weighted graph, and the subgraph strategy is Hybrid TC.\vspace{-6mm}}
    \label{tab:agg-function}
\end{table}

\begin{figure*}
\centering
\subfloat[\small With $\ell_1$]
{\label{fig:dynanom-l1-functions}
 \includegraphics[height=.21\linewidth]{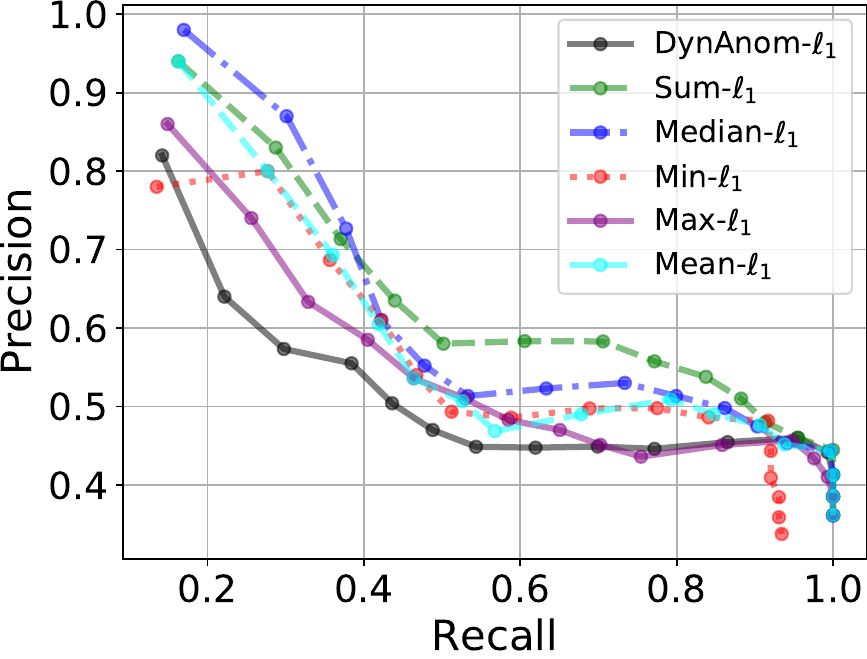}}\
 \subfloat[With $\ell_2$]
{\label{fig:dynanom-l2-functions}
 \includegraphics[height=.21\linewidth]{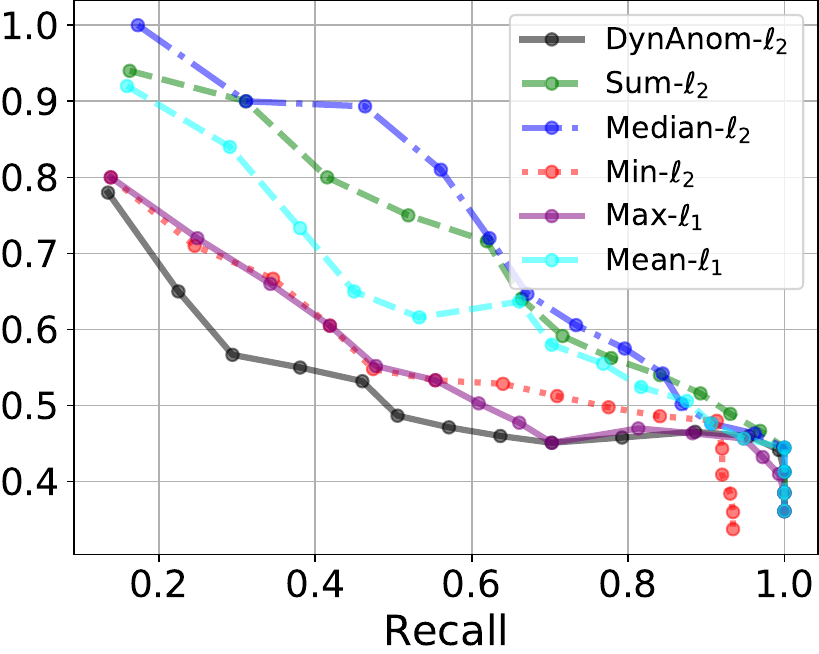}}\
 \subfloat[Pre-Rec curves]
{\label{fig:score_function}
 \includegraphics[height=.21\linewidth]{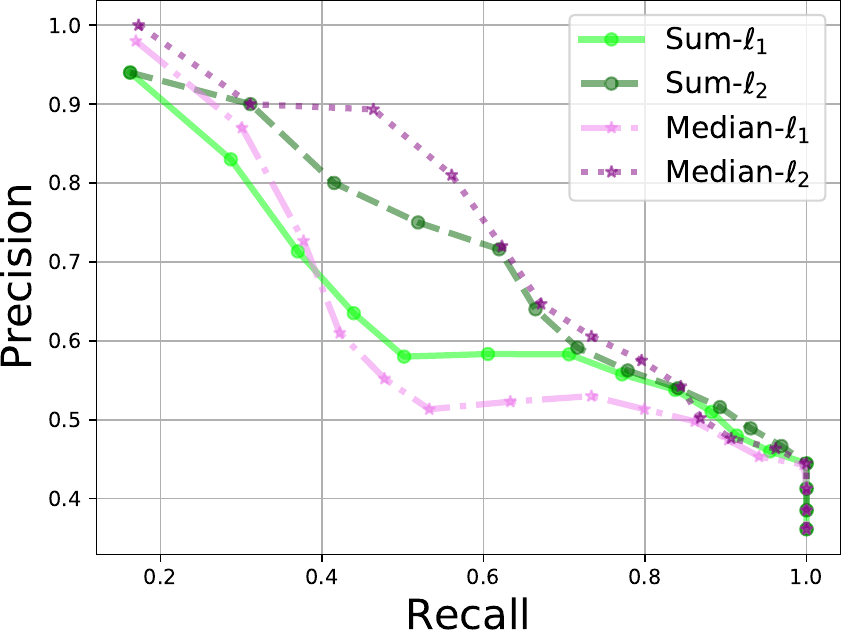}}

\caption{Precision-Recall curves to compare aggregations. \textsc{Median} and \textsc{Sum} perform best, and their performances are close. Given the best subgraph strategy and aggregation functions, $\ell_2$ performs better than $\ell_1$ as the distance function.\vspace{-2mm} }
\label{fig:pre-recall-curve}
\end{figure*}

\begin{figure}
    \centering
    \subfloat[$\ell_1$ on weighted graph] {\includegraphics[height=0.16\textwidth]{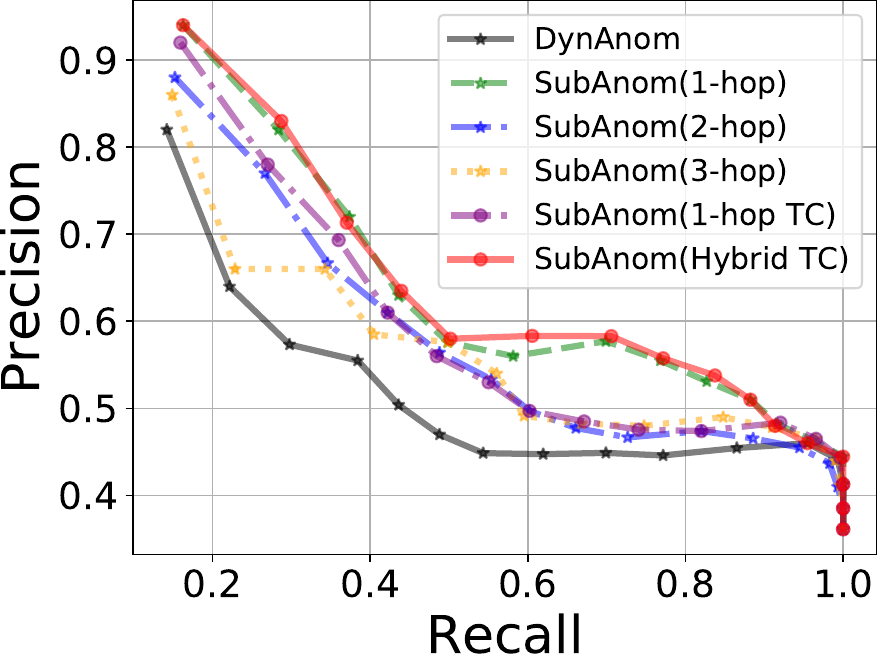}\label{fig:dynanom-l1}}\ 
  \subfloat[$\ell_2$ on weighted graph]
  {\includegraphics[height=0.16\textwidth]{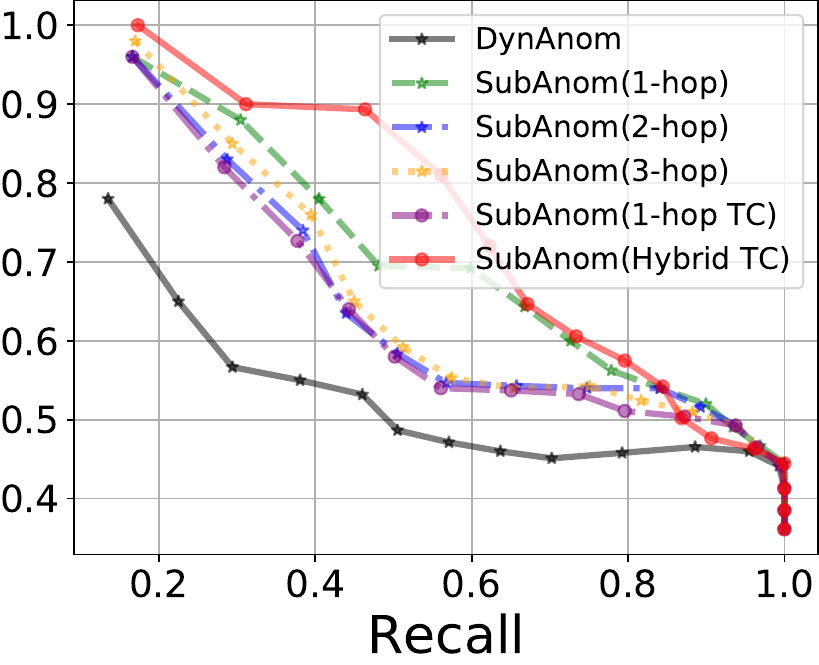}\label{fig:dynanom-l2}} \ 
  \subfloat[$\ell_1$ on unweighted graph]
  {\includegraphics[height=0.16\textwidth]{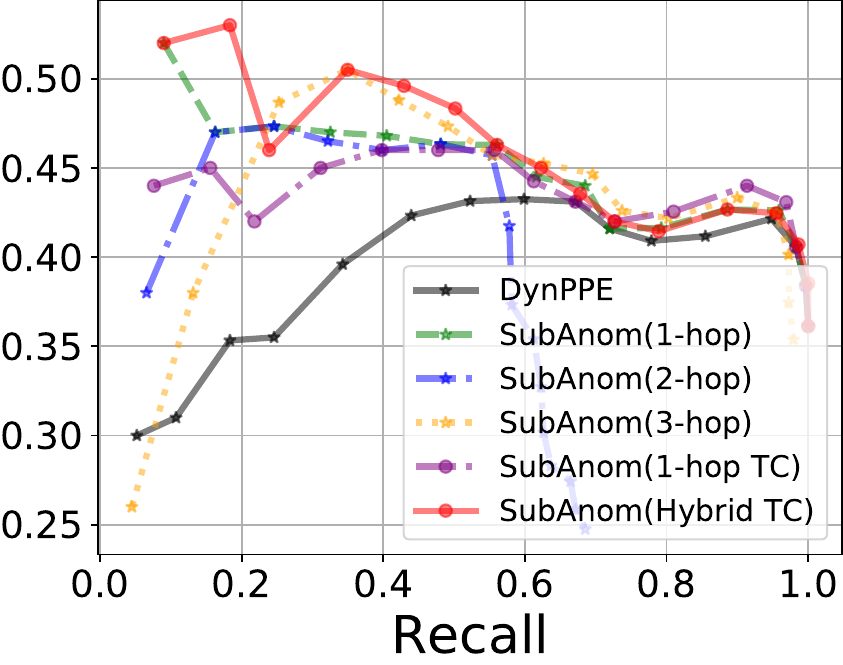}\label{fig:dynppe-l1}}\ 
  \subfloat[$\ell_2$ on unweighted graph]
  {\includegraphics[height=0.16\textwidth]{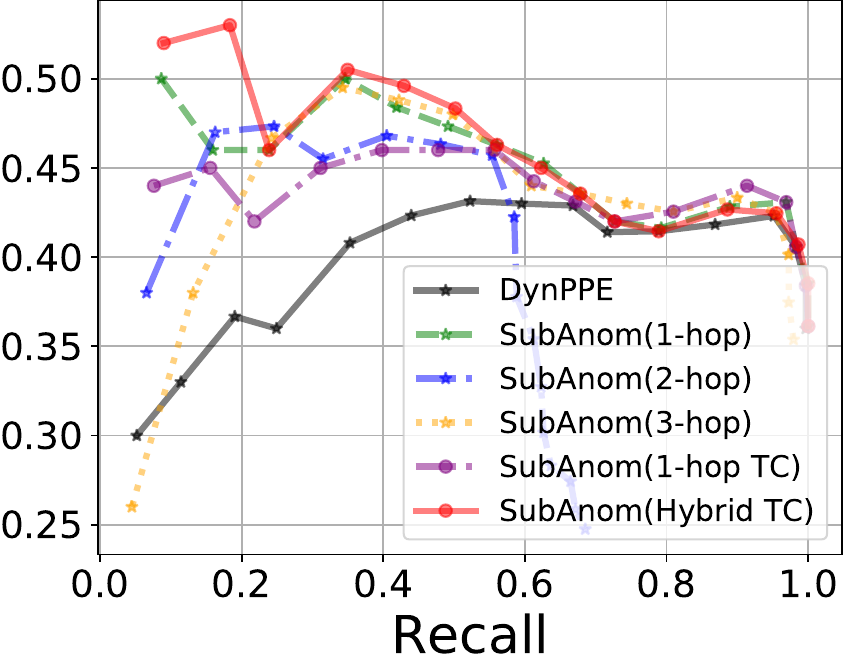}\label{fig:dynppe-l2}}
  \caption{Comparison of subgraph strategies on DARPA dataset\vspace{-8mm}}
  \label{fig:darpa-pr}
\end{figure}

To address Question 1, we leverage a sequence of snapshots, denoted as $\{\G^1,\G^2,\ldots,\G^{1207}\}$. Both \textit{weighted} and \textit{unweighted} dynamic node embeddings are used. Additionally, $\ell_1$ and $\ell_2$ distance metrics are considered. For the subgraph identification strategy, we select $k$-hop with $k=1,2,3$, 1-hop TC, and Hybrid TC. For the aggregation functions of subgraphs, we examine \textsc{Sum}, \textsc{Mean}, \textsc{Median}, \textsc{Max}, and \textsc{Min}. To gauge the detection capability, we identify the top $k'$ highest score snapshots as anomalous graphs, in line with the method suggested by Yoon et al. (2019). We deem 250 detected snapshots as an appropriate benchmark, given that the dataset contains 289 ground-truth anomalies. Fig. \ref{fig:darpa-pr} presents the precision-recall curves as we adjust the parameter $k' \in \{50, 100, ..., 800\}$ for the top-$k'$ snapshots considered anomalous. All subgraph strategies perform better than the baseline method by a large margin, while the performance of the Hybrid TC strategy is the best.

When we employ weighted dynamic node embeddings and use the $\ell_1$-norm as the score function, as shown in Table \ref{tab:darpa-da}, the Hybrid TC strategy emerges as the most effective. Using the \textsc{Sum} operation in the function $\phi$, it delivers the highest precision among the top 250 results, scoring 58.00\%. This precision is 7.60\% higher than that achieved by \textsc{DynAnom} and just 0.4\% lower than \textsc{DynAnom}(High-degree). Figure \ref{fig:dynanom-l1} further illustrates the superior performance of the Hybrid TC strategy. The Precision-Recall curve for Hybrid TC is typically higher than those for the other strategies, suggesting that the boundary nodes produced through Hybrid TC can facilitate improved anomalous subgraph detection.

To testify the robustness of \textsc{SubAnom}, we consider $\ell_2$-norm as the score function, as displayed in Table \ref{tab:darpa-da}, the Hybrid TC strategy again prevails. Adopting the \textsc{Median} operation for $\phi$, Hybrid TC achieves the highest precision among the top 250 results, registering a score of 72.00\%. This precision is 18.80\% higher than \textsc{DynAnom} and 10.80\% higher than \textsc{DynAnom}(High-degree). This is corroborated by the Precision-Recall curve shown in Fig. \ref{fig:dynanom-l2}, which depicts the Hybrid TC strategy outperforming the other strategies in most instances. Thus, the Hybrid TC strategy consistently demonstrates robustness in our evaluations. The Precision-Recall curve of SubAnom($2$-hop) does not perform as effectively as the Hybrid TC strategy. This can be attributed to the fact that as $k$ increases, a larger number of normal nodes are likely to be included, potentially resulting in the omission of anomalous nodes situated at the boundaries.

To provide further insight, we utilized unweighted graph embeddings and both $\ell_1$ and $\ell_2$ as score functions; the results are presented in Table \ref{tab:darpa-dp} and Figure \ref{fig:dynppe-l1}. Among the strategies, the 3-hop strategy achieved the highest precision in the top-250 with \textsc{Median} as $\phi$, reaching a precision of 48.80\%. This precision was 9.20\% higher than that of \textsc{DynPPE} and 2.0\% higher than \textsc{DynPPE}(High-degree). Conversely, the Hybrid TC strategy recorded a precision of 47.20\%, an increase of 7.6\% compared to \textsc{DynPPE} and 0.4\% higher than \textsc{DynPPE}(High-degree). The Precision-Recall curves reveal that although the 3-hop strategy has higher precision within the top 250, the Hybrid TC strategy outperforms it under most conditions. All in all, the Hybrid TC strategy remains the best subgraph strategy across the majority of experimental settings, with \textsc{SubAnom}(Hybrid TC) outperforming baseline methods. 

\subsection{Comparison of Aggregation Methods and Distance Metrics}
In this part, we compare the subgraph aggregation functions and distance metrics. We choose \textsc{DynAnom} as the baseline method, the Hybrid TC strategy as the subgraph strategy, and both $\ell_1$ and $\ell_2$ as distance metrics. We use the Precision-Recall curves and the top 250 snapshots to measure performance. 

\textit{\textbf{Subgraph aggregation functions.}\quad} The results of the best $\phi$ are shown in Fig. \ref{fig:pre-recall-curve} and Table \ref{tab:agg-function}. From Table \ref{tab:agg-function}, \textsc{Median}-$\ell_2$ has the highest F1. From the P-R curves, the curves of \textsc{Sum} and \textsc{Median} are on the top, so \textsc{Sum} and \textsc{Median} are the best subgraph aggregation functions among the functions no matter with $\ell_1$ or $\ell_2$ as the score function, while performances of both $\phi$ are close to each other. 

\textit{\textbf{Embedding distance function.}\quad} We conducted the experiments utilizing the Hybrid TC strategy for subgraph identification and \textsc{Sum} and \textsc{Median} as the $\phi$ aggregation functions. The Precision-Recall (P-R) curves of the experiments are depicted in Fig. \ref{fig:score_function}. The curves for $\ell_2$ are consistently above those for $\ell_1$, which suggests that $\ell_2$ delivers superior performance as a score function. The reason is that the $\ell_2$-norm is generally more sensitive to large differences between pair-wise components, which could make it better suitable for identifying anomalous nodes where large changes have occurred. While both $\ell_1$ and $\ell_2$ norms can be used effectively as score functions in the Hybrid TC strategy, our results suggest that $\ell_2$ provides more accurate identifications.

\section{Discussion and Conclusion}
\label{sect:conclusion}

In this work, we present a novel anomalous subgraph detection framework, \textsc{SubAnom}, that operates at the subgraph level based on dynamic graph embedding by identifying not just anomalous snapshots but also subgraphs. For future work, it would be interesting to extend our approach to dynamic attributed graphs. This extension could provide an even more nuanced understanding of anomalies in network structures.

\section{Acknowledgements}
The authors would like to thank the anonymous reviewers for
their helpful comments. The work of Baojian Zhou is sponsored
by the Shanghai Pujiang Program (No. 22PJ1401300). The work of
Deqing Yang is supported by Chinese NSF Major Research Plan
No.92270121, Shanghai Science and Technology Innovation Action
Plan No.21511100401.
\bibliographystyle{unsrt}
\bibliography{SUBANOM}

\end{document}